\documentclass[twocolumn,amsmath,amssymb]{snp}
\usepackage{graphicx}
\usepackage{dcolumn}
\usepackage{bm}
\def\be {\begin{equation}}
\def\ee {\end{equation}}
\def\nn {\nonumber}
\def\bea {\begin{eqnarray}}
\def\eea {\end{eqnarray}}

\newcommand{\om}{\omega}

\newcommand{\bp}{\boldsymbol{p}}

\begin{document}

\title{{\Large On lower bound of relaxation time for massless fluid in presence of magnetic field}}

\author{\large Sarthak Satapathy$^{1,*}$, Jayanta Dey$^1$, Prashant Murmu$^2$ Sabyasachi Ghosh$^1$}
\email{sarthaks@iitbhilai.ac.in}
\affiliation{$^1$Indian Institute of Technology Bhilai, GEC Campus, Sejbahar, Raipur 492015, 
Chhattisgarh, India}
\affiliation{$^2$ Indian School of Mines Dhanbad 826004, Jharkhand, India}

\maketitle

Present work has gone through analytic calculation of shear viscosity and entropy density of massless relativistic
fluid, facing external magnetic field $B$ and then explored its fluid property.

Before going to finite magnetic field picture, let us
remind the expression of shear viscosity for bosonic/fermionic fluid in absence of magnetic field~\cite{Gavin},
\be
\eta=\frac{\beta}{15}\int \frac{d^3\bp}{(2\pi)^3}\frac{\bp^4}{\om^2}\tau_c f (1-af)~,
\label{eta_B0}
\ee
where $f=1/(e^{\beta\om}+a)$, with $a=\pm 1$ for fermionic/bosonic fluid, $\om=\sqrt{\bp^2+m^2}$ is
energy of massive constituent of medium, and most important quantity is the relaxation time $\tau_c$,
which proportionally control the strength of shear viscosity. 

Now, when we come the magnetic field
picture, an additional time scale $\tau_B=\om/(eB)$ will come into the picture and in terms of
these two time scale $\tau_c$ and $\tau_B$, we will get an effective time scale
\be
\tau_{\rm eff}=\tau_c/\Big\{1+\Big(\frac{\tau_c}{\tau_B}\Big)^2\Big\}
\ee
which will be entered in place of $\tau_c$ of Eq.~(\ref{eta_B0}) for finite magnetic field picture.
Hence we will get shear viscosity of bosonic/fermionic fluid in presence of magnetic field~\cite{JD_eta} as
\be
\eta_2=\frac{\beta}{15}\int \frac{d^3\bp}{(2\pi)^3}\frac{\bp^4}{\om^2}\tau_{\rm eff} f (1-af)~,
\label{eta_B}
\ee
where the subscript $2$ stands for one of the component of shear viscosities among 5 components,
appeared due to magnetic field.

For massless case, Eq.~(\ref{eta_B0}) becomes
\bea
\eta &=& \frac{4\:\tau_c}{5\pi^2}\zeta(4) \: T^4=\frac{4\pi^2\:\tau_c}{450} \: T^4 ~~{\rm for~Boson}
\nn\\
&=& \Big(\frac{7}{8}\Big)\frac{4\:\tau_c}{5\pi^2}\zeta(4) \: T^4=\frac{7\pi^2\:\tau_c}{900} \: T^4~~{\rm for~Fermion}
\nn\\
\label{eta_anl}
\eea
and Eq.~(\ref{eta_B}) becomes
\bea
\eta_2&=& \frac{\eta(B=0)}{1+(\tau_c/\tau_B)^2}=\frac{\frac{4\pi^2\:\tau_c}{450} \: T^4}{1+(\tau_c/\tau_B)^2} ~~{\rm for~Boson}
\nn\\
&=& \frac{\eta(B=0)}{1+(\tau_c/\tau_B)^2}=\frac{\frac{7\pi^2\:\tau_c}{900} \: T^4}{1+(\tau_c/\tau_B)^2}~~{\rm for~Fermion}~.
\nn\\
\label{eta_2}
\eea
For calculation simplification, we have considered average energy in $\tau_B$
\bea
\tau_B&=&\Big\{\frac{\zeta(4)}{\zeta(3)}\Big\}\frac{3T}{eB} ~~{\rm for~Boson}
\nn\\
&=&\Big\{\frac{7\zeta(4)}{2\zeta(3)}\Big\}\frac{3T}{eB} ~~{\rm for~Fermion}
\label{Eav_TB}
\eea 
Now, let us come to the perfect fluid aspects of fermionic/bosonic fluid.
Fluidity of the medium is measured by the shear viscosity to entropy density ratio $\eta/s$,
where entropy density $s$ of medium can be expressed as
\be
s=\int \frac{d^3\bp}{(2\pi)^3}\Big(\om + \frac{\bp^2}{3\om}\Big) f~,
\label{s_B0}
\ee
whose massless limit is
\bea
s &=& \frac{4}{\pi^2}\zeta(4) \: T^3=\frac{4\pi^2}{90} \: T^3 ~~{\rm for~Boson}
\nn\\
&=& \Big(\frac{7}{8}\Big)\frac{4}{\pi^2}\zeta(4) \: T^3=\frac{7\pi^2}{180} \: T^3~~{\rm for~Fermion}~.
\nn\\
\label{s_anl}
\eea
In classical case, we can imagine a perfect fluid, having $\eta/s=0$ but in quantum
case, we get a lower bound of $\eta/s$, which is also known as KSS bound~\cite{KSS}, and it is $1/(4\pi)$.
For massless medium, using $\eta$ from Eq.~(\ref{eta_anl}) and $s$ from Eq.~(\ref{s_anl}), we can get 
$\eta/s=\tau_c T/5$, which is interestingly same for bosonic or fermionic fluid, although their individual $\eta$
and $s$ expressions are different. If at $\tau_c=\tau_c^0$, lower bound of $\eta/s$ for massless medium is
achieved, then we get 
\bea
\frac{\eta}{s}=\frac{\tau^0_c T}{5}&=&\frac{1}{4\pi}
\nn\\
\Rightarrow \tau^0_c &=&\frac{5}{4\pi \: T}~,
\label{tauc0}
\eea
which is quite standard well-known results. 
%
In this context, the present article provide similar type of analytic
expression of relaxation time as a function $T$ and $B$ for massless fluid
in presence of magnetic field.
Using normal viscosity component $\eta_2$ from Eq.~(\ref{eta_2})
and $s$ from Eq.~(\ref{s_anl}), we will get $\eta_2/s$, which will
be now different for Maxwell-Boltzmann (MB), Bose-Einstein (BE) 
and Fermi-Dirac (FD) distribution cases as it contains $\tau_B$, whose
average values are different for different cases.
By restricting $\eta_2/s=1/4\pi$, we can get quadratic equation of $\tau_c$:
\bea
\tau_c^2 - 4\pi \tau_B^2\frac{\eta(B=0)}{s}+\tau_B^2&=&0
\nn\\
\Rightarrow\tau_c^2 - \Big(\frac{4\pi T \tau_B^2}{5}\Big)\tau_c+\tau_B^2&=&0
\nn\\
\Rightarrow\tau_c^2 - \Big(\frac{ \tau_B^2}{\tau^0_c}\Big)\tau_c+\tau_B^2&=&0~.
\eea
The solution of above equation is
\be
\tau_c=\tau_c^{2\pm}=\frac{\tau_B^2}{2\tau^0_c}\Big[1\pm\sqrt{1-4\Big(\frac{\tau^0_c}{\tau_B}\Big)^2}\Big]~.
\label{tauc2}
\ee
So far from our best knowledge, we are first time addressing an {\it analytic} expressions 
of $\tau_c(T,B)$~\cite{JD_eta},
where massless bosonic/fermionic matter in presence of magnetic field reach the KSS bound.
To get a physical solution of Eq.~(\ref{tauc2}), we need
\bea
1-4\Big(\frac{\tau^0_c}{\tau_B}\Big)^2 &\geq &0
\nn\\
\Rightarrow \tau_B &\geq & 2\tau_c^0~.
\eea
Using MB relation $\tau_B=\frac{3T}{eB}$ in above inequality, we have
\bea
\frac{3T}{eB} &\geq & 2\frac{5}{4\pi T}
\nn\\
T &\geq & \Big(\frac{5 eB}{6\pi}\Big)^{1/2}~.
\label{MB_TeB}
\eea
Corresponding FD and BE relations from Eq.~(\ref{Eav_TB}) will give 
\bea
T &\geq & \Big[\Big(\frac{\zeta(3)}{\zeta(4)}\Big)\frac{5 eB}{6\pi}\Big]^{1/2}~{\rm for ~BE}
\nn\\
T &\geq & \Big[\Big(\frac{2\zeta(3)}{7\zeta(4)}\Big)\frac{5 eB}{6\pi}\Big]^{1/2}~{\rm for ~FD}~.
\label{FDBE_TeB}
\eea
Hence, Eqs.~(\ref{MB_TeB}) and (\ref{FDBE_TeB})
can be identified as upper allowed domain, where KSS bound can be achieved while lower domain is
forbidden zone if we believe that $\eta_2/s$ never goes below $\frac{1}{4\pi}$.
Within the allowed $T$-$B$ zone, we will get two positive values of relaxation time $\tau_c^{\pm}$,
where KSS bound can be achieved. It can be understandable mathematically as follows.
For smaller values of $\tau_c$ or $\tau_c/\tau_B<<1$,
$\eta_2/s\propto \tau_c$, while for larger values of $\tau_c$ or $\tau_c/\tau_B>>1$,
$\eta_2/s\propto 1/\tau_c$. Due to this two opposite trends of $\tau_c$ dependence, we
are getting a lower and upper values of $\tau_c$, where $\eta_2/s$ reach the lower bound $1/(4\pi)$.
The detail analysis of this investigation is documented in Ref.~\cite{JD_eta}.

\newpage
\end{document}